\documentclass[superscriptaddress,prd,aps,10pt,twocolumn,floats,nofootinbib,showpacs]{revtex4-1}

\usepackage{amssymb,amsmath}
\usepackage{epsfig}
\pdfoutput=1
\usepackage{amssymb,amsmath}
\usepackage{epsfig}
\usepackage{xcolor}
\usepackage[utf8]{inputenc}
\usepackage[normalem]{ulem}

\newcommand{\be}{\begin{equation}}
\newcommand{\ee}{\end{equation}}
\newcommand{\ba}{\begin{eqnarray}}
\newcommand{\ea}{\end{eqnarray}}
\newcommand{\beg}{\begin{gather*}}
\newcommand{\eng}{\end{gather*}}
\newcommand{\hh}{,\hspace{0.5cm}}
\newcommand{\hhh}{,\hspace{0.2cm}}

\newcommand{\eq}[1]{(\ref{#1})}

\newcommand{\n}[1]{\label{#1}}

\newcommand{\ins}[1]{{\mbox{\tiny #1}}}

\newcommand{\ts}[1]{{\boldsymbol{#1}}}

\newcommand{\wk}{\omega_{\ts{k}}}
\newcommand{\ak}{\hat{a}_{\ts{k}}}
\newcommand{\an}{\hat{a}_{\ts{k}}^{\dagger}}
\newcommand{\akp}{\hat{a}_{\ts{k}'}}
\newcommand{\anp}{\hat{a}_{\ts{k}'}^{\dagger}}
\newcommand{\bk}{\hat{b}_{\ts{k}}}
\newcommand{\bn}{\hat{b}_{\ts{k}}^{\dagger}}
\newcommand{\bkp}{\hat{b}_{\ts{k}'}}
\newcommand{\bnp}{\hat{b}_{\ts{k}'}^{\dagger}}
\newcommand{\Fk}{\Phi_{\ts{k}}(\ts{x})}
\newcommand{\CFk}{\Phi_{\ts{k}}^{*}(\ts{x})}

\def\XXint#1#2#3{{\setbox0=\hbox{$#1{#2#3}{\int}$ }
\vcenter{\hbox{$#2#3$ }}\kern-.6\wd0}}

\usepackage[makeroom]{cancel}
\usepackage[caption=false]{subfig}
\usepackage[colorlinks=true,
            citecolor=green,
            linkcolor=red,
            filecolor=cyan,
            urlcolor=magenta,
            backref=false]{hyperref}

\newcommand{\dd}{\mbox{d}}

\begin{document}

\title{Resonant particle creation by a time-dependent potential in a nonlocal theory}
\author{Jens Boos}
\email{jboos@wm.edu}
\affiliation{Theoretical Physics Institute, University of Alberta, Edmonton, Alberta, Canada T6G 2E1}
\affiliation{High Energy Theory Group, Department of Physics, William \& Mary, Williamsburg, VA 23187-8795, United States}
\author{Valeri P. Frolov}
\email{vfrolov@ualberta.ca}
\affiliation{Theoretical Physics Institute, University of Alberta, Edmonton, Alberta, Canada T6G 2E1}
\author{Andrei Zelnikov}
\email{zelnikov@ualberta.ca}
\affiliation{Theoretical Physics Institute, University of Alberta, Edmonton, Alberta, Canada T6G 2E1}

\date{\today}

\begin{abstract}
Considering an exactly solvable local quantum theory of a scalar field interacting with a $\delta$-shaped time-dependent potential we calculate the Bogoliubov coefficients analytically and determine the spectrum of created particles. We then show how these considerations, when suitably generalized to a specific nonlocal ``infinite-derivative'' quantum theory, are impacted by the presence of nonlocality. In this model, nonlocality leads to a significant resonant amplification of certain modes, leaving its imprint not only in the particle spectrum but also in the total number density of created particles.
\end{abstract}


\maketitle

\section{Introduction}

The effect of quantum particle creation in time-dependent backgrounds has many interesting applications in all branches of physics. In particular, it provides a mechanism for the production of density fluctuations during the inflationary stage of the Early Universe \cite{Chibisov:1981wa,Mukhanov:1981xt,Mukhanov:1990me}, giving rise not only to the present large-scale structures in the Universe but also to the measured cosmic microwave background anisotropies. It is widely expected that gravity theory is to be modified by suitable UV completions at least at the Planck scale, though experimentally this modification is not ruled out for scales less than 50 $\mu$m \cite{Lee:2020zjt}. The existence of a new fundamental scale $\ell$ can have important implications for our understanding of the Early Universe. In the literature there are many different modifications of gravity involving a minimal length scale, see Ref.~\cite{Hossenfelder:2012jw} for a review.

Here we consider a particular class of \emph{nonlocal} generalizations of relativistic field theories wherein the scale of nonlocality is encoded in a Lorentz invariant form factor depending on the \emph{scale of nonlocality} $\ell$; these theories have been studied quite extensively \cite{Modesto:2010uh,Biswas:2011ar,Frolov:2015bia,Carone:2016eyp,Modesto:2017ycz,Koshelev:2018hpt}. Such form factors improve the UV behavior of theories \cite{Talaganis:2014ida,Pius:2016jsl,Carone:2016eyp,Briscese:2018oyx} without introducing spurious degrees of freedom, which is why this type of theories are called ghost-free. Nonlocal form factors naturally appear in the context of quantum gravity and have been explored in the context of cosmology \cite{Biswas:2010zk,Biswas:2012bp,Briscese:2012ys,Briscese:2013lna,Koshelev:2016xqb,Koshelev:2020xby} as well as black hole singularities \cite{Biswas:2011ar,Frolov:2015usa,Koshelev:2017bxd}.

Here we study imprints of nonlocality on the quantum effect of particle creation. While nonlocality complicates computations of quantum effects considerably, it is nevertheless possible to address some questions by considering an exactly solvable model. Exact solutions provide us with analytical results for scattering amplitudes, and allow the analysis of non-perturbative and non-analytical aspects of these problems. The results obtained for such an idealized model reveal nevertheless robust qualitative effects one can expect in more realistic systems.

A well known problem, both in the non-relativistic and the relativistic case, is the quantum-mechanical scattering of a particle on a $\delta(x)$-like potential. The scattering problem on such a potential in the framework of a nonlocal scalar field theory was recently studied in Ref.~\cite{Boos:2018kir}. Beyond this simple model, in General Relativity the gravitational perturbations on the background of an expanding FLRW universe obey an effective scalar field equation with a time-dependent potential. For this reason a scalar field theory is a good starting point to study the evolution of quantum perturbations of gravitational fields. In the nonlocal case, however, the situation is more complicated and requires further analysis.

In this Letter we consider the effect of particle creation from vacuum via a time-dependent $\delta(t)$-like potential in a nonlocal scalar field theory. Our results provide some intuition of what one can expect in a more realistic setup: We show that nonlocality can lead to a significant resonant amplification of particle creation for some wavelengths defined by the scale of nonlocality $\ell$ and the potential strength $\lambda$. This effect is absent for the local theory with the same $\delta$-potential, hence providing an interesting new effect solely due to the presence of nonlocality.

\section{Particle creation by a time-dependent potential}

\subsection{Free fields}

Consider $D=(d+1)$-dimensional Minkowski spacetime and let $X{}^\mu = (t,\ts{x})$ be its Cartesian coordinates in which the metric reads
\begin{align}
\dd s^2 = \eta{}_{\mu\nu} \dd X{}^\mu \dd X{}^\nu = -\dd t^2 + \dd \ts{x}^2 \, .
\end{align}
Consider first a local free massless scalar  quantum field $\hat{\varphi}(X)$ obeying the equation
\begin{align}\n{BB}
\Box \, \hat{\varphi}(X)=0\, .
\end{align}
We write this field in the form
\begin{align}\n{SS}
\hat{\varphi}(t,\ts{x})=\int {\dd^d k\over \sqrt{2\omega_{\ts{k}}}} \left[ \hat{a}_{\ts{k}} e^{-i\wk t} \Fk+\an e^{i\wk t}\CFk\right] \, .
\end{align}
Here we denote $\wk=|\ts{k}|$, the spatial basis as
\begin{align}
\Fk={1\over (2\pi)^{d/2} } e^{i\ts{k}\cdot\ts{x}}\, ,
\end{align}
and $\ak$ and $\an$ are annihilation and creation operators, respectively, obeying the canonical commutation relations
\begin{align}
[\ak,\anp]=\delta{}^{(d)}(\ts{k}-\ts{k}')\hhh
[\ak,\akp]=[\an,\anp]=0\, .
\end{align}
The vacuum state $|0\rangle$, which describes a state with no positive frequency particles, is defined by the following condition [see also Eq.~\eq{akbk} in appendix \ref{app}]:
\begin{align}
\ak |0\rangle =0.
\end{align}
Let us discuss now a nonlocal theory of the scalar massless field which is obtained by substituting the $\Box$-operator in \eqref{BB} by the operator
\be \n{DD}
{\cal D}=f(\ell^2\Box) \Box\, .
\ee
The function $f$ which enters this expression is called a form factor and $\ell>0$ is the \emph{scale of nonlocality}.
We assume that $f(z)$ does not have zeroes in the complex plane of the complex variable $z$. In this case, the inverse of the operator \eqref{DD} does not introduce new poles besides those of the $\Box$-operator, indicating that such a theory does not have new spurious degrees of freedom (ghosts). To construct such a theory it is sufficient to choose $f(z)=\exp[g(z)]$, where $g(z)$ is an entire function. For example, one can take $g(z)$ to be a polynomial, and in order to facilitate analytical calculations we shall use the form factor $f(z)=\exp(z^2)$.

In the absence of sources and external potentials, the nonlocal field equation
\be \n{NL}
{\cal D}\hat{\varphi}(X)=0
\ee
has the same solutions as the local equation \eqref{BB}, which can be written in the form \eqref{SS}, implying that on-shell there is no difference between the models (\ref{BB}) and (\ref{NL}).

\subsection{Time-dependent potential: local case}
Let us consider now a local theory of a scalar field with a time-dependent potential $V(t)$,
\begin{align}\n{PHI-local}
\big[\Box -V(t)\big]\hat{\varphi}(X)=0\, .
\end{align}
Its solutions can be expanded in modes
\begin{align}\n{TD}
\hat{\varphi}(t,\ts{x})=\int \dd^d k \left[ \hat{\varphi}_{\ts{k}}(t) \Fk+\hat{\varphi}_{\ts{k}}^{\dagger}(t)\CFk\right] \, .
\end{align}
In the local case each mode $ \hat{\varphi}_{\ts{k}}(t)$ obeys the equation
\begin{align}\n{OM}
[\partial_t^2+\Omega_{\ts{k}}^2(t)] \hat{\varphi}_{\ts{k}}(t)=0 \, ,
\end{align}
and $\Omega_{\ts{k}}^2(t)=\omega_{\ts{k}}^2+V(t)$. Let us assume that the potential $V(t)$ vanishes or becomes very small outside some time interval. In these past and future time domains the frequency $\Omega_{\ts{k}}(t)$ is constant and coincides with $\wk$.

Denote by ${\varphi}_{\ts{k}}(t)$ a complex solution of the equation
\begin{align} \n{PHI}
[\partial_t^2+\Omega_{\ts{k}}^2(t)] \varphi_{\ts{k}}(t)=0\, ,
\end{align}
which has the distant past and future asymptotics
\begin{align}
\label{PHI-asymptotic}
\varphi_{\ts{k}}\big|_{t\to \mp\infty}= \begin{cases} \displaystyle {1\over  \sqrt{2\omega_{\ts{k}}}}  e^{-i\wk t} \equiv \varphi_\ts{k}^0(t) \, , \\[10pt]
\displaystyle {1\over\sqrt{2\omega_{\ts{k}}}} \left( \alpha_{\wk} e^{-i\wk t}
+\beta_{\wk} e^{i\wk t} \right) \, . \end{cases}
\end{align}
This choice of $\varphi_\ts{k}^0(t)$ corresponds to the standard scattering problem. It describes the setup of the system with only positive frequency modes in the distant the past. The complex coefficients $\alpha_{\wk}$ and $\beta_{\wk}$ are called Bogoliubov coefficients and can be obtained by solving Eq.~\eqref{PHI}. Let us emphasize that these coefficients depend on the frequency $\wk$ rather than wave vector $\ts{k}$, making them invariant under the reflection $\ts{k}\to -\ts{k}$. The Bogoliubov coefficients satisfy [see Eq.~\eq{ab} in appendix \ref{app}]
\begin{align}\n{BAC}
|\alpha_{\wk}|^2-|\beta_{\wk}|^2=1\, .
\end{align}
Denoting the number of particles in the mode $\ts{k}$ created from vacuum by the time-dependent potential as $n_{\wk}$ and the total density of particles as $N$ one has
\begin{align} \n{NNa}
&n_{\wk}=|\beta_{\wk}|^2 ,\\
&N = \int \dd^d k \, n_{\wk} = {2\pi^{d/2}\over \Gamma\left(\tfrac{d}{2}\right)}\int \dd\omega \,\omega^{d-1} n_{\omega} \, .
\end{align}
Equation \eqref{PHI} coincides with that of a harmonic oscillator with a time-dependent frequency. If the corresponding quantum oscillator is initially in its ground state, as a result of such parametric excitations it can jump into excited levels, where the expression \eqref{NNa} describes the probability of such a process.

\subsection{Time-dependent potential: nonlocal case}

The nonlocal field equation is of the form
\begin{align}\n{PHI-nonlocal0}
\big[\mathcal{D} -V(t)\big]\hat{\varphi}(X)=0\hh \mathcal{D} = \exp\left(\ell^4\Box^2\right) \, \Box \, ,
\end{align}
where $\ell$ is the scale of nonlocality. We assume again that the potential $V(t)$ vanishes or becomes very small outside some time interval. The effects of nonlocality are controlled by the length scale parameter $\ell$, such that in the remote past and remote future the asymptotic solutions of Eq.~\eqref{PHI-nonlocal0} coincide with the solutions of the local equation. Hence one can define creation and annihilation operators in these past and future asymptotic domains exactly in the same way as in the local theory. The only difference is that the vacuum state can be defined uniquely only in the asymptotic region, where the notion of a particle is understood. In these regions one can then employ the standard local creation and annihilation operator algebra, such that in order to find a relation between them one can proceed as in the local case. Namely, one can write a solution in the form \eqref{TD} where now the equation for $\varphi_\ts{k}(t)$-modes is
\begin{align}\n{PHI-nonlocal}
\left\{ e^{\ell^4(\partial_t^2+\wk^2)^2} [\partial_t^2+\omega_\ts{k}^2 ] + V(t) \right\} \varphi_\ts{k}(t) = 0 \, .
\end{align}
By solving this nonlocal equation one can extract the coefficients $ \alpha_{\wk}$ and $\beta_{\wk}$ via Eq.~\eqref{PHI-asymptotic} by relating the in- and out-asymptotics of the solutions, and thereby calculate the number of particles in the mode $\ts{k}$ created by the time-dependent potential in the nonlocal model.

\section{Exactly solvable model}
\subsection{Local case}
It is possible to determine the Bogoliubov coefficients analytically in some cases. Let us first demonstrate this in the local theory. In the absence of the time-dependent potential, the retarded Green function $G^\text{R}_\ts{k}(t'-t)$ solves
\begin{align}
[\partial_t^2+\wk^2] G^\text{R}_\ts{k}(t'-t) = -\delta(t'-t) \, ,
\end{align}
while also satisfying $ G^\text{R}_\ts{k}(t'-t) \equiv 0$ if $t'<t$. One finds
\begin{align}
G^\text{R}_\ts{k}(t'-t) =- \frac{\sin[\wk(t'-t)]}{\wk}\theta(t'-t) \, .
\end{align}
One can use it to obtain the Lippmann--Schwinger integral representation \cite{Lippmann:1950zz} for the solution of Eq.~\eqref{PHI},
\begin{align}
\varphi_\ts{k}(t) = \varphi_\ts{k}^0(t) + \int\limits_{-\infty}^\infty \dd t' G^\text{R}_\ts{k}(t-t') V(t')\varphi_\ts{k}(t') \, .
\end{align}
Here, $\varphi_\ts{k}^0(t)$ is a free solution which satisfies $[\partial_t^2+\wk^2]\varphi_\ts{k}^0(t)=0$, and we may choose it to correspond to the asymptotic past of Eq.~\eqref{PHI-asymptotic}.\footnote{This choice of $\varphi_\ts{k}^0(t)$ corresponds to the proper ``in'' vacuum state, and is also ideally suited to discuss ghost-free non-local theories, since this choice only demands a certain asymptotic behavior.} In the special case of a $\delta$-shaped potential, $V(t) = \lambda\delta(t)$, this integral collapses:
\begin{align}
\varphi_\ts{k}(t) = \varphi_\ts{k}^0(t) + \lambda\varphi_\ts{k}(0)G^\text{R}_\ts{k}(t) \, ,
\end{align}
where $ \varphi_\ts{k}^0(t)= e^{-i\wk t}/\sqrt{2\omega_{\ts{k}}}$.
It is solved by
\begin{align}
\varphi_\ts{k}(t) = \varphi_\ts{k}^0(t) + \frac{\lambda}{\sqrt{2\wk}} G^\text{R}_\ts{k}(t) \, .
\end{align}
Using Eq.~\eqref{PHI-asymptotic} we can read off the Bogoliubov coefficients
\begin{align}
\alpha_{\wk} = 1 - \frac{i\lambda}{2\wk} \, , \quad \beta_{\wk} = \frac{i\lambda}{2\wk} \, .
\end{align}
The spectral density and the total number density of particles are given by the expressions
\begin{align}
n_{\wk} = \frac{\lambda^2}{4\wk^2} \, , \quad N =  {\pi^{d/2}\over 2\Gamma\left(\tfrac{d}{2}\right)}\lambda^2 \int\limits_0^\infty \dd k\,k^{d-3}\, .
\end{align}
For $d\ge2$ the number density of created particles $N$ formally diverges at large momenta $k$, which is due to the infinitely small width of the $\delta$-potential. For potentials of finite time duration $\tau$, there appears an effective cut-off $k_\ins{max}\sim 1/\tau$ at high momenta \cite{Dunne:1998ni} and the number density of created particles is indeed finite.

\subsection{Nonlocal case}
Let us now consider a nonlocal model and study how the presence of nonlocality affects the Bogoliubov coefficients and, ultimately, the particle spectrum as well as total number density of produced particles. In the limiting case of $\ell\rightarrow 0$ one recovers the local theory, but for $\ell>0$ the above differential operator can lead to nonlocal behavior \cite{Boos:2019fbu,Boos:2019zml,Boos:2019vcz,Boos:2020qgg}. The corresponding nonlocal retarded Green function, in the absence of the potential, is a solution of
\begin{align}
e^{\ell^4(\partial_t^2+\wk^2)^2} [\partial_t^2+\wk^2] \mathcal{G}^\text{R}_\ts{k}(t'-t) = -\delta(t'-t) \, .
\end{align}
It can be decomposed as
\begin{align}\label{DeltaG}
&\mathcal{G}^\text{R}_\ts{k}(t'-t) = G^\text{R}_\ts{k}(t'-t) + \Delta\mathcal{G}_\ts{k}(t'-t) \, , \\
\label{DeltaG-2}
&\Delta\mathcal{G}_\ts{k}(t) = \int\limits_{-\infty}^\infty \frac{\dd \omega}{2\pi} e^{-i\omega t} \frac{e^{-\ell^4(\omega^2-\wk^2)^2} - 1}{\omega^2-\wk^2} \, .
\end{align}
It has been shown that at late and early times $\Delta\mathcal{G}_\ts{k}$ vanishes and one has \cite{Boos:2019zml}
\begin{align}
\label{eq:deltaG-asymptotic}
\mathcal{G}_\ts{k}\big|_{t\to \pm\infty} = G^\text{R}_\ts{k}(t) \, .
\end{align}
Using the nonlocal retarded Green function we can express a solution of Eq.~\eqref{PHI-nonlocal0} as
\begin{align}
\varphi_\ts{k}(t) = \varphi_\ts{k}^0(t) + \int\limits_{-\infty}^\infty \dd t' \mathcal{G}^\text{R}_\ts{k}(t-t') V(t')\varphi_\ts{k}(t') \, .
\end{align}
Here, $\varphi_\ts{k}^0(t)$ is again a free solution, and due to \eqref{eq:deltaG-asymptotic} we may choose it identical to the local case. Recall that it describes the solution with only positive frequency modes in the distant past. This is an important point because we are interested in particle creation from vacuum. One may consider also other solutions and study amplification or damping of various modes but they would correspond to non-vacuum initial states.

For the $\delta$-shaped potential we then find
\begin{align}
\varphi_\ts{k}(t) = \varphi_\ts{k}^0(t) + \lambda\varphi_\ts{k}(0) \mathcal{G}^\text{R}_\ts{k}(t) \, .
\end{align}
Now substitute here $t=0$ and solve the resulting algebraic equation for $\varphi_\ts{k}(0)$ in terms of $ \varphi_\ts{k}^0(0)$, yielding
\begin{align}
\varphi_\ts{k}(t) = \varphi_\ts{k}^0(t) + \frac{\lambda}{1-\lambda \mathcal{G}^\text{R}_\ts{k}(0)}\frac{1}{\sqrt{2\wk}} \mathcal{G}^\text{R}_\ts{k}(t) \, .
\end{align}
Using the future asymptotic Eq.~\eqref{PHI-asymptotic} as well as \eqref{eq:deltaG-asymptotic} we can read off the Bogoliubov coefficients at $t\rightarrow\infty$,
\begin{align}
\alpha_{\wk} &= 1 - \frac{i\lambda}{2\wk}\frac{1}{1-\lambda\Delta\mathcal{G}_\ts{k}(0)} \, , \\
\beta_{\wk} &= \frac{i\lambda}{2\wk}\frac{1}{1-\lambda\Delta\mathcal{G}_\ts{k}(0)} \, .
\end{align}
The number of particles in the mode $\ts{k}$ created by the time-dependent potential is given by
\begin{align} \n{NNb}
n_{\wk}=|\beta_{\wk}|^2=\frac{\lambda^2}{4\wk^2\big|1-\lambda\Delta\mathcal{G}_\ts{k}(0)\big|^2}\, .
\end{align}
Recall that, as in the local case, the total number of created particles diverges at large momenta. However, the integral over the momenta is truncated at high frequencies $k_\ins{max}\sim 1/\tau$ if the potential has a finite width (duration of time) $\tau$. However, if one solely considers the nonlocal contribution to the particle number density the UV cutoff problem does not appear at all.

\subsection{Resonant particle creation}

\begin{figure}
    \centering
	\vspace{1.5em}
    \includegraphics[width=0.45\textwidth]{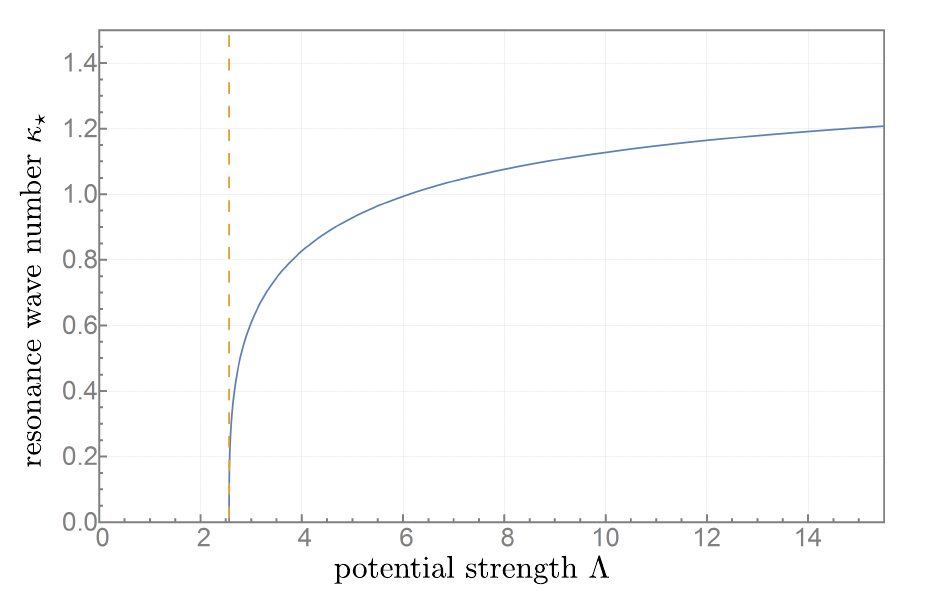}
    \caption{The resonance wave number $\kappa_\star = |\ts{k}_\star|\ell$  as a function of $\Lambda=\lambda\ell$. The resonant amplification occurs if the potential strength exceeds the critical value $\Lambda_\ins{crit} \ge \pi/\Gamma\left(\tfrac{3}{4}\right)$. }
    \label{kappastar}
\end{figure}

As a new nonlocal effect, the denominator in \eqref{NNb} may vanish. At this wavelength some kind of resonant particle creation takes place, leading to a huge amplification of modes with the wave vector $\ts{k}_\star$ that satisfies the condition
\begin{align}\label{DGk}
\Delta\mathcal{G}_\ts{k_\star}(0)=\frac{1}{\lambda} \, .
\end{align}
This condition defines a resonant frequency $\omega_\star=|\ts{k}_\star|$. Inserting $t=0$ into Eq.~\eqref{DeltaG-2} and following the derivation in \cite{Boos:2019fbu} one can show
\begin{align}
\begin{split}
\label{eq:gf2-deltaG}
\Delta\mathcal{G}_\ts{k}(0) &= \frac{\Gamma\left(\tfrac34\right)\ell}{\pi} {}_\ins{2}F{}_\ins{2}\left( \tfrac14, \tfrac34; \tfrac12,\tfrac54; -\kappa^4 \right) \\
&\hspace{11pt} - \frac{\sqrt{2}\kappa^2\ell}{6\Gamma\left(\tfrac34\right)} {}_\ins{2}F{}_\ins{2}\,\left( \tfrac34, \tfrac54; \tfrac32,\tfrac74; -\kappa^4\right) \, ,
\end{split}
\end{align}
where ${}_\ins{2}F{}_\ins{2}$ is the generalized hypergeometric function \cite{Olver:2010}. We also defined the dimensionless wave number $\kappa = k\ell$. Introducing the dimensionless coupling $\Lambda = \lambda\ell$, the resonance condition \eq{DGk} takes the simple form $\mathcal{F}= 1/\Lambda$ with $\mathcal{F} = \Delta\mathcal{G}_\ts{k}(0)/\ell$. It has a solution $\kappa_\star$ provided
\begin{align}
\Lambda\ge \Lambda_\ins{crit}= \frac{\pi}{\Gamma\left(\tfrac{3}{4}\right)} \approx 2.56369\dots \, ,
\end{align}
see also Fig.~\ref{kappastar}. The function $\mathcal{F}$ has the asymptotics
\begin{align}
\mathcal{F} \approx \begin{cases} \displaystyle \frac{\Gamma\left(\tfrac34\right)}{\pi}- \frac{\sqrt{2}\kappa^2}{6\Gamma\left(\tfrac34\right)} & \text{ for } \kappa \rightarrow 0 \, , \\[10pt]
\displaystyle \frac{1}{4\sqrt{\pi}\kappa^3} & \text{ for } \kappa \rightarrow \infty \, . \end{cases}
\end{align}
At very high frequencies, the nonlocal modification decreases and hence the density of created particles asymptotically approaches the local theory for those high frequencies. For $\Lambda \to \Lambda_\text{crit}$ the resonant nonlocal amplification happens at
\begin{align}
\kappa_\star\approx
\left[\frac{6\Gamma\left(\tfrac34\right)}{\sqrt{2}}
\Big(\frac{\Gamma\left(\tfrac34\right)}{\pi}-\frac{1}{\Lambda}\Big)\right]^{1/2}\, .
\end{align}
When $\Lambda \to \infty$ it corresponds to
\begin{align}
\kappa_\star\approx\Big(\frac{\Lambda}{4\sqrt{\pi}}\Big)^{1/3}
\hh
k_\star\approx\Big(\frac{1}{4\sqrt{\pi}}\frac{\lambda}{\ell^2}\Big)^{1/3}\, .
\end{align}

While the local particle spectrum is invariant under the transformation $\lambda \rightarrow -\lambda$, this is no longer the case in the nonlocal theory. In fact, the resonant amplification occurs solely for positive $\lambda\ge\Lambda_\ins{crit}/\ell$. At the level of the local Bogoliubov coefficients, a transformation $\lambda\rightarrow-\lambda$ merely induces a phase shift of $\pi$ in the far future modes. This is no longer true in the nonlocal theory, and could be related to the fact that nonlocality smears the phases of incoming plane waves in vicinity of the source.

For wave numbers close to the resonance $k_\star$ the density of created particles \eq{NNb} diverges as $\sim (k-k_\star)^{-2}$ and their integral number diverges. This can be the result of idealization of the potential as a $\delta$-function, which necessarily includes infinitely high wave numbers. In a realistic system the potential has a finite width, and for such a potential of the characteristic width $\tau$ only the modes with $k<k_\ins{max}\sim 1/\tau$ are created, which means that one can expect resonant particle number amplification for potentials satisfying the approximate condition $(\lambda\ell)^{-1/3} < \tau/\ell < 1$. It is natural to expect for realistic localized potentials that the width of the resonant band will be finite and the total number density of created particles also will be regularized.

\section{Discussion}
We computed the rate of scalar particle creation in the framework of ghost-free nonlocal field theory in the presence of a time-dependent potential. Our particular model of nonlocality was dictated by absence of instabilities in this model \cite{Frolov:2016xhq} as well as because of its analytically known Green function, allowing for an exact study of the problem including non-perturbative effects.

We found a new, nonlocal effect: there appears a new feature of resonant particle creation at some frequency $\omega_\star=\kappa_\star/\ell$. This effect depends on the sign and the amplitude $\lambda$ of the $\delta(t)$-potential. In the limit when the scale of nonlocality $\ell$ tends to zero, we naturally recover the predictions for particle creation at all frequencies except a very narrow peak at high frequencies $\omega_\star$, which grows with decreasing $\ell$. In other words, in the local limit the resonant frequency is shifted to infinity.

We believe that qualitative effects found in our exactly solvable model are robust and are applicable to a wide variety of more realistic nonlocal theories. An interesting question is: Can the nonlocal parametric resonant amplification effect, discussed in this Letter, manifest itself as a potentially observable imprint on the spectrum of the primordial perturbations in early inflationary cosmology and be tested in observations? We shall leave this question for future work.

\section*{Acknowledgments}

We thank our anonymous referee for spotting a sign error in the local retarded Green function. J.B.\ is grateful for a Vanier Canada Graduate Scholarship administered by the Natural Sciences and Engineering Research Council of Canada as well as for the Golden Bell Jar Graduate Scholarship in Physics by the University of Alberta, and was supported in part by the National Science Foundation under grant PHY-1819575. V.F.\ and A.Z.\ thank the Natural Sciences and Engineering Research Council of Canada and the Killam Trust for their financial support.

\appendix
\section{Bogoliubov transformation}
\label{app}

Here we briefly summarize the main conceptual steps relating the in-vacuum and out-vacuum in the framework of Bogoliubov coefficients. In the distant future domain one has
\begin{align}\n{aa}
\hat{\varphi}(t,\ts{x})&=\int {\dd^d k\over \sqrt{2\omega_{\ts{k}}}} \left[
(\alpha_{\wk}\hat{a}_{\ts{k}} +\beta_{\wk}^{*}\hat{a}^{\dagger}_{-\ts{k} })e^{-i\wk t} \Fk  \right. \nonumber \\
&\hspace{12pt}+ \left. (\alpha_{\wk}^{*}\an + \beta_{\wk}\hat{a}_{-\ts{k} }) e^{i\wk t}\CFk\right] \, .
\end{align}
On the other hand, one has
\begin{align} \n{bb}
\hat{\varphi}(t,\ts{x})=\int {\dd^d k\over \sqrt{2\omega_{\ts{k}}}} \left[ \hat{b}_{\ts{k}} e^{-i\wk t} \Fk+\bn e^{i\wk t}\CFk\right] \, ,
\end{align}
where $\bk$ and $\bn$ are the annihilation and creation operators of out-particles, respectively. By comparing (\ref{aa}) and (\ref{bb}) one obtains
\be\n{BBK}
\bk=\alpha_{\wk}\hat{a}_{\ts{k}} +\beta_{\ts{k}}^{*}\hat{a}^{\dagger}_{-\ts{k} } \, .
\ee
This constitutes a Bogoliubov transformation connecting in- and out-particles. Since the operators $\bk$ and $\bn$ obey the commutation relations
\be
[\bk,\bnp]=\delta(\ts{k}-\ts{k}')\hhh
[\bk,\bkp]=[\bn,\bnp]=0\, ,
\ee
the Bogoliubov coefficients satisfy the relation
\be\label{ab}
|\alpha_{\wk}|^2-|\beta_{\wk}|^2=1\, .
\ee
If the Bogoliubov coefficients are known one can construct a unitary $\hat{S}$-matrix relating in- and out-states of the system \cite{berezin}. In this sense they encode the complete information about the system. The in- and out-vacuum states are defined by conditions
\be\label{akbk}
\ak |0; \text{in}\rangle =0\hh \bk |0; \text{out}\rangle =0\, .
\ee
To simplify notations we omit the index ``in'' for in-states. For example, $|0 \rangle$ denotes the in-vacuum state.
The in-state with $m_{\ts{k}}$ particles is
\be
 |m_{\ts{k}}\rangle ={(\an)^{m_{\ts{k}}}\over \sqrt{m_{\ts{k}} !}} |0 \rangle \, .
\ee
These states have the property
\be
\langle m_{\ts{k}'} ' |m_{\ts{k}}\rangle =\delta_{m_{\ts{k}},m_{\ts{k}} '} \delta^{(d)}(\ts{k}-\ts{k}')\, .
\ee
Using (\ref{BBK}) it is easy to show that
\be
\bk |m_{\ts{k}}\rangle =\alpha_{\wk} \sqrt{m_{\ts{k}}} |m_{\ts{k}}-1\rangle
+\beta_{\wk} \sqrt{m_k+1}|m_{\ts{k}}+1\rangle\, .
\ee
This relation implies that the number of out-particles $n_{\ts{k}}$ created from the initial state with $m_{\ts{k}}$ particles is
\be
n_{\ts{k}}=m_{\ts{k}}+(m_{\ts{k}}+1)|\beta_{\wk}|^2\, .
\ee
If the system is in the initial vacuum state, then the number of out-particles created in the mode $\ts{k}$ from this state by a time-dependent field is
\be \n{NN}
n_{\wk}=\langle 0|\bn \bk |0\rangle =|\beta_{\wk}|^2\, .
\ee
Finally, the total number $N$ of particles created by the time-dependent potential is given by the integral
\be
N=\int \dd^d k \, n_{\wk}={2\pi^{d/2}\over \Gamma(d/2)}\int \dd\omega \omega^{d-1} n_{\omega}\, .
\ee

\bibliography{Ghost_references}{}


\end{document}